\documentclass[preprint,showpacs,prb,floatfix,superscriptaddress,citeautoscript
,cite]{revtex4}
\usepackage{graphicx}
\usepackage{color}
\usepackage{amsmath}

\begin{document}

\title{Bilayer SnS$_{2}$: Easy-tunable Stacking Sequence by Charging and 
Loading Pressure}

\author{C. Bacaksiz}
\affiliation{Department of Physics, Izmir Institute of Technology, 35430, Izmir, 
Turkey}

\author{S. Cahangirov}
\affiliation{UNAM-National Nanotechnology Research Center, Bilkent University, 
06800 Ankara, Turkey}

\author{A. Rubio}
\affiliation{Nano-Bio Spectroscopy Group and ETSF, Dpto. Fisica de Materiales, 
Universidad del Pais Vasco, CFM CSIC-UPV/EHU-MPC and DIPC, 20018 San Sebastian, 
Spain}
\affiliation{Max Planck Institute for the Structure and Dynamics of Matter and 
Center for Free-Electron Laser Science, Luruper Chaussee 149, 22761 Hamburg, 
Germany}

\author{R. T. Senger}
\affiliation{Department of Physics, Izmir Institute of Technology, 35430, Izmir, 
Turkey}

\author{F. M. Peeters}
\affiliation{Department of Physics, University of Antwerp, Groenenborgerlaan 
171, 2020 Antwerp, Belgium}

\author{H. Sahin}
\affiliation{Department of Physics, University of Antwerp, Groenenborgerlaan 
171, 2020 Antwerp, Belgium}

\date{\today}

\begin{abstract}

Employing density functional theory-based methods, we investigate monolayer and 
bilayer structures of hexagonal SnS$_{2}$, which is recently synthesized 
monolayer 
metal dichalcogenide. Comparison of 1H and 1T phases of monolayer SnS$_{2}$ 
confirms the ground state to be the 1T phase. In its bilayer structure we 
examine different stacking configurations of the two layers. It is found that 
the interlayer coupling in bilayer SnS$_{2}$ is weaker than that of typical 
transition-metal dichalcogenides (TMDs) so that alternative stacking orders have 
similar structural parameters and they are separated with low energy barriers. 
Possible signature of the stacking order in SnS$_{2}$ bilayer has been sought in 
the calculated absorbance and reflectivity spectra. We also study the effects of 
the external electric field, charging, and loading pressure on the 
characteristic properties of bilayer SnS$_{2}$. It is found that (i) the 
electric field increases the coupling between the layers at its prefered 
stacking order, so the barrier height increases, (ii) the bang gap value can be 
tuned by the external E-field and under sufficient E-field, the bilayer 
SnS$_{2}$ can become semi-metal, (iii) the most favorable stacking order can be 
switched by charging and (iv) a loading pressure exceeding 3 GPa changes the 
stacking order. E-field tunable bandgap and easy-tunable stacking sequence of 
SnS$_{2}$ layers make this 2D crystal structure a good candidate for field 
effect transistor and nanoscale lubricant applications. 
\end{abstract}

\maketitle

\section{Introduction}

Ultrathin materials,\cite{Novoselov1} the study of which was boosted after 
the synthesis of graphene,\cite{Novoselov2} have attracted considerable 
interest 
due to their remarkable physical properties.\cite{Butler,Chhowalla} Graphene has 
extraordinary mechanical\cite{Lee} and optical\cite{Nair} properties. However, 
due to the lack of a band gap in graphene,\cite{Zhou} exploring other 
two-dimensional (2D) materials with a band gap became important for several 
applications. In this respect, synthesis and theoretical prediction of many 
other 2D materials have been achieved, such as  silicene,\cite{Cahangirov,Kara} 
germanene,\cite{Cahangirov,Ni,Davila,Yang} stanene,\cite{Guzman,Bechstedt} 
transition metal dichalcogenides 
(TMDs),\cite{Gordon,Coleman,Wang1,Ross,Sahin2,Tongay,Horzum,Chen3} and 
hexagonal structures of III-V binary compounds (e.g. h-BN, 
h-AlN).\cite{Sahin3,Wang2,Kim,Tsipas,Bacaksiz} The atomic scale of thickness of 
these materials led to new physical insights which 
suggests that possible other 2D materials may exhibit novel properties. In 
addition, the need for a wide range of materials for device technology makes the 
discovery of new  layered materials essential.

In regard to search for new graphene-like or TMD-like 2D material, 
Sn-dichalcogenides are good candidate because of their vdW-linked lamellar 
crystal structure and energy bandgap which is in the visible frequency region. 
As a member of this family, tin disulfide (SnS$_{2}$) was previously 
investigated in the bulk form for various 
applications.\cite{Fotouhi,Parkinson1,Parkinson2,Delawski,Schlaf,Panda,Ma}
After the emergence of novel 2D materials and improved production methods such 
as chemical vapor deposition, chemical and mechanical exfoliation, thinner 
structures of SnS$_{2}$ were synthesized for different applications. For 
example, a few nanometer-thick hexagonal SnS$_{2}$ was used for lithium storage 
in battery applications.\cite{Kim2,Seo,Zhai,Ma2} To enhance the electrochemical 
performance, composite forms of SnS$_{2}$ with graphene were 
examined.\cite{Jiang,Zhuo,Qu,Zhou2,Huang1} Single- and few-layer SnS$_{2}$ were 
also used to fabricate a field-effect transistor.\cite{Pan,De,Song} Moreover, 
photocatalytic character of single- and few-layer SnS$_{2}$ was shown in 
different studies which is directly related to the optical properties of 
hexagonal SnS$_{2}$.\cite{Chen,Chao,Sun,Wei} Furthermore, SnS$_{2}$ nanosheet 
was studied for photosensitive field emission and photodetector 
applications.\cite{Joshi,Xia}

Recently, Zhang \textit{et al.} demonstrated that photoluminescence spectrum of SnS$_{2}$ and MoS$_{2}$ show  additional features when they form a van der Waals heterostructure which is important for the engineering of their 
electronic and optical properties.\cite{Zhang2} Huang \textit{et al.} 
investigated the synthesis, characterization and the electronic properties of 
SnS$_{2}$, from bulk to monolayer.\cite{Huang} More recently, Su \textit{et al.} 
reported that hexagonal SnS$_{2}$ is a suitable material for photodetection 
applications with fast photocurrent response time $\sim$5 $\mu$s.\cite{Su} In 
addition to these, Ahn \textit{et al.} successfully synthesized hexagonal 
SnS$_{2}$ and orthorhombic SnS as a polymorphic 2D heterostructure.\cite{Ahn}

Although there are a few number of computational works on single layer 
hexagonal SnS$_{2}$\cite{Zhuang2,Xia2}, comprehensive investigation of 
electronic and optical properties of it's monolayer and bilayer crystal 
structures are still lacking. Therefore, in this study we concentrate on the 
monolayer and bilayer forms of  hexagonal SnS$_{2}$. The structural parameters, 
electronic properties and optical response of these materials are investigated 
using ab initio methods. In addition, from the calculated absorbance or 
reflectivity spectra, the optical signatures which allows one to characterize 
the structural phase or the stacking order of the SnS$_{2}$ layers were sought. 
Furthermore, we investigate the effects of an applied perpendicular electric 
field, charging, and loading pressure on the characteristic properties of 
bilayer SnS$_{2}$.

The paper is organized as follows: in Sec. \ref{comp} we give details of our 
computational methodology. An overview of the structural phases, the electronic 
and optical properties of monolayer hexagonal SnS$_{2}$ are presented in Sec. 
\ref{monolayer}. In Sec. \ref{bilayer} different stacking orders of bilayer 
SnS$_{2}$ in the T-phase are investigated in detail. The effect of the external 
electric field, charging, and loading pressure on the bilayer system are 
studied. Finally, we present our conclusion in Sec. 
\ref{conc}.

\section{Computational Methodology}\label{comp}

Our investigations of the structural, electronic and optical properties for the 
layered SnS$_{2}$ were performed using the Vienna ab-initio simulation package, 
VASP\cite{vasp1,vasp2,vasp3} which is based on density functional theory 
(DFT). The VASP code solves the Kohn-Sham equations for a 
system with periodic boundary conditions using iteratively a plane-wave basis 
set. The Perdew-Burke-Ernzerhof (PBE) form of the generalized gradient 
approximation (GGA)\cite{GGA-PBE} was adopted to describe electron 
exchange and correlation. The hybrid DFT-HSE06 functional\cite{HSE06} on top of 
GGA was used for a more accurate estimation of the band gap, as compared to GGA 
which usually underestimates the band gap of semiconducting systems. The 
spin-orbit interaction, which is essential for the TMDs, was included in the 
calculations. The interlayer interaction is dominated by the vdW forces for 
such layered materials, which was taken into account by using the DFT-D2 
method of Grimme.\cite{vdW1,vdW2} To obtain the charge distribution of the
configuration, a Bader charge analysis is used.\cite{Bader1,Bader2}

Structural optimizations were performed with the following parameters. The 
kinetic energy cut-off of the plane-wave basis set was 500 eV in all 
calculations. The total energy difference between the sequential steps in the
iterations was taken 10$^{-5}$ units as convergence criterion. The convergence 
for the Hellmann-Feynman forces on each atom was taken to be 10$^{-4}$ 
eV/\AA{}. Gaussian smearing of 0.05 eV was used and the pressures on the unit 
cell were decreased to a value less then 1.0 kB in all three directions. For the
determination of accurate charge densities, Brillouin zone integration was 
performed using a $35\times35\times1$ $\Gamma$-centered mesh for the primitive 
unit cell. To avoid interactions between adjacent SnS$_{2}$ monolayers and few 
layer systems, our calculations were performed with a large unit cell including 
∼16 \AA{} vacuum space.

\begin{figure}
\includegraphics[width=8.5 cm]{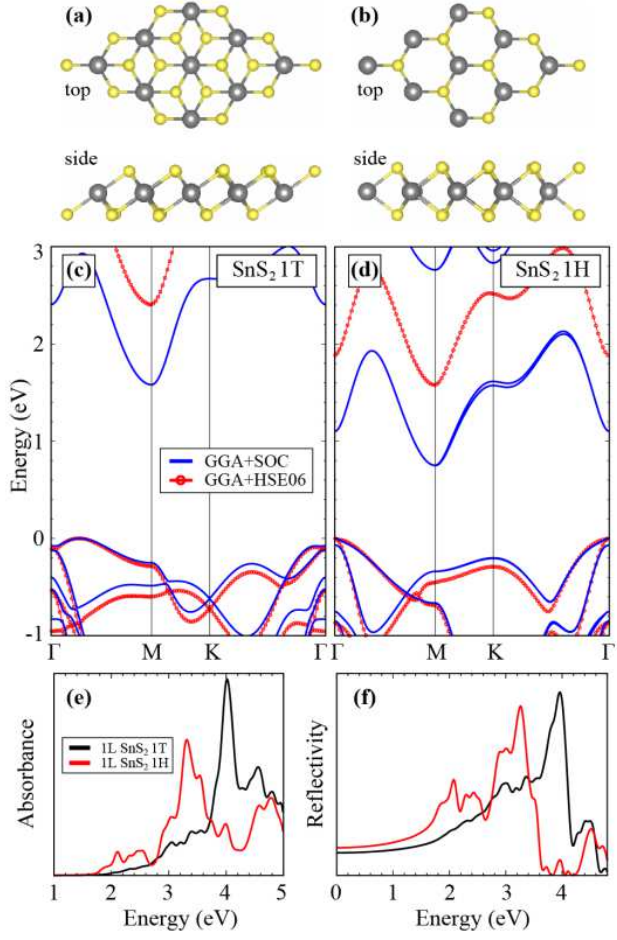}
\caption{\label{1lfig}
(Color online) (a) and (b) illustrate the structure of monolayer 1T and 1H 
SnS$_{2}$, respectively. (c) and (d) are the band structures of 1T and 1H 
SnS$_{2}$. The blue curves and the curve with red-circles are 
for GGA+SOC and GGA+HSE06, respectively. (e) and (f) are absorbance and the 
reflectivity of the 1T (black 
curve) and 1H SnS$_{2}$ (red curve).}
\end{figure}

In addition, the absorbance and the reflectivity of the previously optimized 
structures was investigated with and without spin-orbit interaction and also 
including HSE06 on top of GGA. A $\Gamma$-centered $k$-point sampling of 
$70\times70\times1$ was used for monolayer systems. Because of the computational 
burden, $k$-point sampling was reduced to $35\times35\times1$ for the bilayer 
systems, and $21\times21\times1$ for the calculations that include the HSE06 
hybrid functional and spin-orbit interaction. The calculated dielectric function 
provides us with the optical quantities such as the frequency dependent 
absorbance $A(\omega)$ and the Fresnel reflectivity $R(\omega)$ through the 
formulas;
\begin{align}
A(\omega) &= \frac{\omega}{c} L \textbf{Im}[\epsilon(\omega)] \\
 \nonumber \\ 
R(\omega) &= \Bigg|\frac{\sqrt{\epsilon(\omega)}+1}{\sqrt{\epsilon(\omega)}-1} \Bigg|^{2}
\end{align}where the dielectric function is defined as $\epsilon(\omega) = 
\epsilon_{1}(\omega) + i \epsilon_{2}(\omega)$ and $\omega$ is the frequency, 
$c$ is the speed of light, $L$ is the unitcell length in the perpendicular 
direction,

\begin{table}
\caption{\label{1ltable} Calculated parameters for monolayer SnS$_{2}$ are the 
lattice constant in the lateral direction, $a$; the distance between the 
subplanes of S, $c$; the intralayer atomic distance, d$_{Sn-S}$; the charge 
transfer from Sn 
to S atom, $\Delta\rho$; the work function $\Phi$; and the cohesive energy, 
E$_{c}$. E$_{g}^{\text{GGA}}$ and E$_{g}^{\text{HSE06}}$ are the 
energy band gap values within GGA+SOC and GGA+HSE06, respectively.}
\begin{tabular}{rcccccccc}
\hline\hline
 & a   & c & d$_{Sn-S}$ & $\Delta\rho$ &$\Phi$& E$_{c}$ & E$_{g}^{\text{GGA}}$ 
& 
E$_{g}^{\text{HSE06}}$\\
 & (\AA) & (\AA) & (\AA) & ($e^{-}$) &(eV)& (eV) & (eV) & (eV) \\
\hline
1T SnS$_{2}$ & 3.68 & 2.96 & 2.59  & 0.7 &7.53& 3.79 & 1.58 & 2.40 \\
1H SnS$_{2}$ & 3.60 & 3.23 & 2.63  & 0.7 &6.19& 3.49 & 0.78 & 1.58 \\
\hline\hline
\end{tabular}
\end{table}

\section{H and T Phases of Single Layer SnS$_{2}$}\label{monolayer}

Monolayer SnS$_{2}$ possesses two different phases 1T and 1H as shown in 
Fig.~\ref{1lfig}. Both phases have three trigonal subplanes where the Sn 
subplane is sandwiched by two S-subplanes. The 1T phase is a member of the 
$P\overline{3}m2$ space group where subplanes of it are $ABC$ stacked. The 1H 
is a member of the $P\overline{6}m2$ space group where subplanes of it are 
$ABA$ stacked. The lattice vectors of both phases are 
$\textbf{v}_{1}=a(\frac{1}{2},\frac{\sqrt{3}}{2},0)$,
$\textbf{v}_{2}=a(\frac{1}{2},-\frac{\sqrt{3}}{2},0)$ where
$|\textbf{v}_{1}|=|\textbf{v}_{2}|$ and $a$ is the lattice constant. The atomic 
coordinates of 1T phase are $(\frac{|v_{1}|}{2},\frac{|v_{1}|}{2},0)$, 
$(\frac{|v_{1}|}{6},\frac{|v_{1}|}{6},\frac{c}{2})$, and 
$(\frac{5|v_{1}|}{6},\frac{5|v_{1}|}{6},-\frac{c}{2})$ for the Sn atom and the 
S atoms, respectively, where $c$ is the distance between the subplanes of S 
atoms. The atomic coordinates of 1H phase are given as 
$(\frac{|v_{1}|}{3},\frac{|v_{1}|}{3},0)$, 
$(\frac{2|v_{1}|}{3},\frac{2|v_{1}|}{3},\frac{c}{2})$, and 
$(\frac{2|v_{1}|}{3},\frac{2|v_{1}|}{3},-\frac{c}{2})$. 

We obtained the lattice constants of 3.68 \AA{} and 3.60 \AA{} for 1T and 1H, 
respectively. The corresponding Sn-S bond lengths (d$_{Sn-S}$) are 2.59 \AA{} 
and 2.63 \AA{} which are given in Table \ref{1ltable}. The energy difference 
between the 1T and 1H phases is $875$ meV per unit cell which shows that the 
formation of the 1H phase is less favorable than 1T. The cohesive energies of 
1T and 1H phases are 3.79 eV and 3.49 eV, respectively. These results are 
consistent with the previous results which find the 1T phase the most favorable 
form of the monolayer. In addition, the work functions ($\Phi$) of the phases 
are 7.54 eV and 6.19 eV. These work function values are larger than those of 
graphene and bilayer graphene ($\sim$4.6 and $\sim$4.7 eV,\cite{Yu} 
respectively) and of single- and few-layer MoS$_{2}$ ($\sim$5.4 eV 
\cite{Choi}). 

Band structures of 1T and 1H phases based on GGA including 
spin-orbit coupling (SOC) and HSE06 hybrid functional are given in Fig. 
\ref{1lfig}. The 1T phase of SnS$_{2}$ monolayer has an indirect band gap where 
the valance band maximum (VBM) is between the $\Gamma$ and $M$ points and the 
conduction band minimum (CBM) is at the $M$ point. As given in Table 
\ref{1ltable}, the band gap of 1T phase is 1.58 eV within GGA+SOC and 2.40 eV 
within GGA+HSE06. 0.7 $e^{-}$ is donated to each S atom by Sn atom. The 1H 
phase also has an indirect band gap where the VBM is at $\Gamma$ point and the 
CBM is at $M$ point. The band gap values are 0.78 eV within GGA+SOC and 1.58 eV 
within GGA+HSE06.

The effect of the SOC is evident in both the 1T and 1H structures, as shown in 
Figs.~\ref{1lfig} (c) and (d). In the 1T structure, the splitting is $\sim$50 
meV at the highest VB states at the $\Gamma$ point while in the 1H structure 
the splittings are $\sim$69 meV in the highest VB states at the $\Gamma$ point 
and $\sim$43 meV in the lowest CB states at the $K$ point. These splittings can 
be exploited in `valleytronics' applications where the excitations of the 
electrons with  different spin are controlled by the polarization of the 
incident light. This was recently demonstrated for the TMDs, especially for 
MoS$_{2}$.\cite{Mak,Cao}

The absorbances and the reflectivities of monolayer SnS$_{2}$ are also 
calculated and the energy dependent plots are given in Figs. \ref{1lfig} (e) 
and (f), respectively. The absorbance plot shows that the 1T and 1H phases have 
different characters. For the 1T phase, absorbance (black) starts at $\sim$1.8 
eV, and at around 4 eV a peak is found. It has also a local maximum at around 
4.5 eV. On the other hand, for the 1H phase the absorbance (red) starts at 
$\sim$1.7 eV and it shows its main peak around 3.2 eV, a local maximum around 
4.8 eV. Since the absorbance spectra of the alternative phases 
are quite distinguishable, optical absorbance measurements can be a reliable tool for 
determining the structural phase of monolayer SnS$_{2}$ samples.

\begin{figure}[htbp]
\includegraphics[width=8.5 cm]{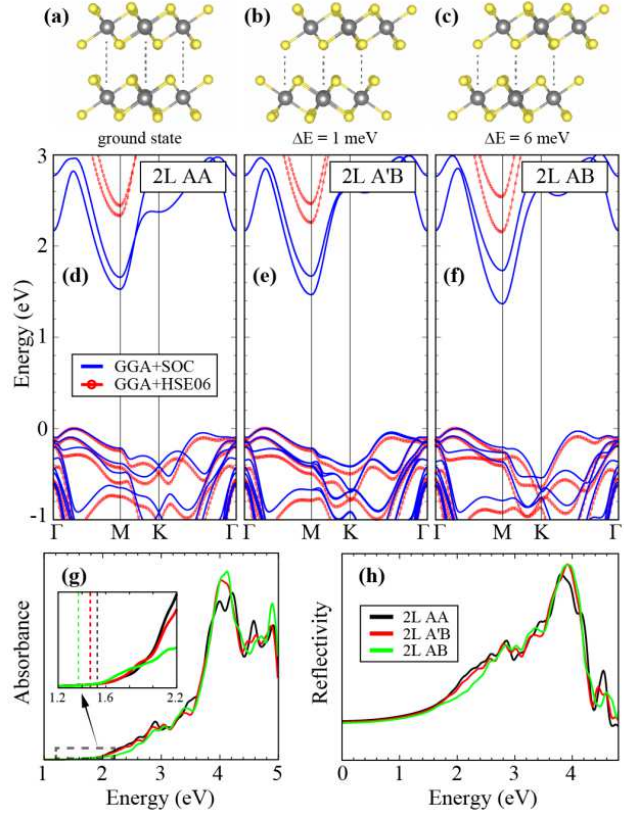}
\caption{\label{2lfig}
(Color online) (a), (b), and (c) are side views of bilayer SnS$_{2}$, and (d), 
(e), and (f) are band structures within the GGA+SOC (blue) and GGA+HSE06 
(red-circle) for $AA$, $A^{\prime}B$, and $AB$ stackings, respectively. (g) and 
(h) are the calculated absorbance and reflectivity of the $AA$ (black), 
$A^{\prime}B$ (red), and $AB$ (green). The vertical lines in the inset of (g) 
represent the band gap values of the corresponding stacking orders.}
\end{figure}

\section{Bilayer SnS$_{2}$}\label{bilayer}

Determining or controlling the stacking order of a layered material is 
important for electronic and optical applications. They can modify the 
electronic and the optical properties even if the layers are weakly interacting 
as in van der Waals layered materials. Improvements in synthesis techniques 
allows researchers to control the stacking order of multilayer structures and 
synthesis of devices with desired features. Therefore,  in this section we 
investigate properties of bilayer SnS$_{2}$ starting with the analysis of 
possible stacking orders.

\begin{table}[htbp]
\caption{\label{2ltable} Calculated values for possible stacking types of 
bilayer 1T-SnS$_{2}$ of the lattice constant in the lateral 
direction, $a$; the distance between the S sublayers of the layers, d$_{L-L}$; 
the energy difference between the structures per SnS$_{2}$, $\Delta$E; 
interlayer interaction potential per formula, E$_{L-L}$; the work function, 
$\Phi$; and the cohesive energy, E$_{c}$. E$_{g}^{\text{GGA}}$ and 
E$_{g}^{\text{HSE06}}$ are the energy band gap values 
within GGA+SOC and GGA+HSE06, respectively.}
\begin{tabular}{rcccccccc}
\hline\hline
    &a& d$_{L-L}$& $\Delta$E & E$_{L-L}$ &$\Phi$& E$_{c}$ &E$_{g}^{\text{GGA}}$ 
& E$_{g}^{\text{HSE06}}$\\
    & (\AA) &   (\AA)  & (meV) & (meV)  &(eV)& (eV)  &  (eV)  &  (eV)  \\
\hline
$AA$          & 3.68 & 2.95 &  0   & 38 &6.50& 3.81  & 1.53  &  2.34    \\
$A^{\prime}B$ & 3.68 & 2.97 &  1   & 38 &6.50& 3.81  & 1.47  &  2.27    \\
$AB$          & 3.68 & 3.03 &  6   & 35 &6.58& 3.81  & 1.37  &  2.17    \\
\hline\hline
\end{tabular}
\end{table}

In Fig.~\ref{2lfig}, bilayer
structures with three different stacking types, their corresponding band 
diagrams, and the imaginary part of the dielectric functions are given. In the 
monolayer section, the 1T phase was found to be energetically favorable, and 
therefore we restrict ourselves to 1T phase. $AA$ (Sn atoms are aligned on Sn 
atoms), $AB$ (S atoms are aligned on Sn atoms) and $A^{\prime}B$ (similar with 
$AB$ but the bottom layer is upside-down) are considered. Also the 
$AA^{\prime}$ stacking (not shown) where the S atoms are aligned on S is 
examined, but its total energy is considerably larger as compared to the given 
three other stacking types. A few meV energy difference was found between the 
$AA$, $A^{\prime}B$, and $AB$ stacking orders which are given in Table 
\ref{2ltable} where we have set the minimum energy to 0. All types have the 
same lattice constant of 3.68 \AA{}. The interlayer distances are 2.95 \AA{}, 
2.97 \AA{}, and 3.03 \AA{} for the $AA$, $A^{\prime}B$, and $AB$ stacking 
orders, respectively.

The cohesive energy of bilayer SnS$_{2}$ in all stacking 
orders are the same, 3.81 eV which is slightly higher than the monolayer 1T 
phase. The work functions are also similar where the values are 6.50 eV, 6.50 
eV and 6.58 eV for the $AA$, $A^{\prime}B$, and $AB$, respectively. The work 
function of the bilayer is smaller than that of the 1T monolayer which is in 
contrast with what was found for graphene and MoS$_{2}$.\cite{Yu,Choi} Another 
point is that the interlayer potential energy per SnS$_{2}$ for the 
different stacking types are also very close to each other; $38$ meV, $38$ 
meV, and $35$ meV for $AA$, $A^{\prime}B$, and $AB$, respectively. This weak 
interaction is a characteristic feature of van Waals layered materials, yet 
these energy values are smaller as compared to graphite (30-55 
meV per atom)\cite{Liu2,Chen2} and typical
TMDs (74, 107, 90, 126 meV per MX$_{2}$ for MoS$_{2}$, MoSe$_{2}$, WS$_{2}$, 
WSe$_{2}$, respectively).\cite{He}

\begin{figure}[htbp]
\includegraphics[width=8.5 cm]{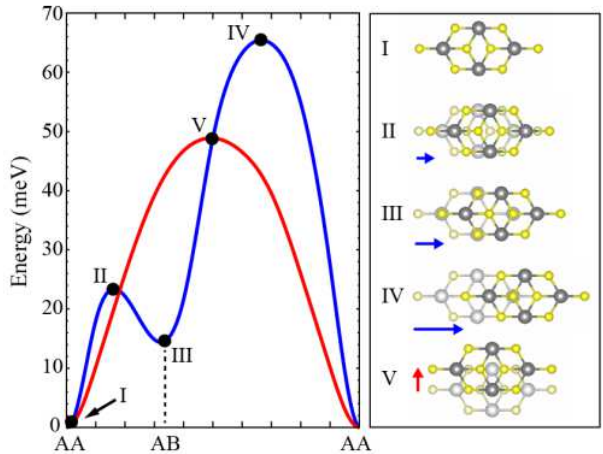}
\caption{\label{sliding}
(Color online)  Left panel, blue (red) curve indicates the energy difference 
when the top layer is sliding along the armchair (zig-zag) direction. Right 
panel, the structural forms of the extremas marked on the energy profiles are 
given.}
\end{figure}

In spite 
of the weak interlayer interactions, and the similarities of the 
structural parameters, the $AA$, $AB$ and $A^{\prime}B$ stacking 
types possess different band dispersions and band gaps. Although the 
VBM and 
the CBM are at the same symmetry points for all stacking orders, the values of 
the indirect band gaps are different. For $AA$, which is energetically the 
favorable one, we have 1.53 eV band gap within GGA and 2.34 eV within HSE06. 
The band gaps in the $A^{\prime}B$ and the $AB$ stackings are 1.47 and 1.37 eV 
within GGA and 2.27 and 2.17 eV within HSE06, respectively. It must be 
emphasized that the band dispersions arising from each bilayer configuration 
differ especially at the symmetry points $M$ and $K$ which are 
important for the optical transitions and the excitonic states. For $AA$ 
stacking, the two CB edge states at the $M$ point are very close to each other 
as compared to those of $A^{\prime}B$ and $AB$. On the other hand, the two CB 
edge states at the $K$ point are significantly different in energy as compared 
to those of $A^{\prime}B$ and $AB$. The absorbance spectrum of the bilayer 
systems are given in Figs.~\ref{2lfig} (g) and (h). The general trend of the 
absorbances for all bilayers are similar. Inset of Fig.~\ref{2lfig} (g)is a 
zoom at the onset region of the absorbance spectrum. Despite the weak 
interactions given in Table \ref{2ltable}, the absorbance spectra provide 
information on the stacking. In addition, the main peak around 4 eV of the $AA$ 
stacking displays two distinct peaks while $A^{\prime}B$ and $AB$ have only one 
peak. Hence, the simple absorbance spectrum carries structural signatures 
although the structures are energetically very similar.

The weak layer-layer interaction in bilayer SnS$_{2}$ is also promising for 
barrierless sliding applications. The sliding  potential in the armchair and 
the zig-zag directions are given in Fig.~\ref{sliding}. The local and global 
extrema and the corresponding structural forms are shown. The positions of the 
upper S atoms of the bottom layer and lower S atoms of the top layer are 
responsible for the potential profile. In the case of sliding along the 
armchair direction the local maximum is seen when the lower S atoms of the top 
layer are positioned at the mid point of the upper S atoms of the bottom layer. 
This is followed by a local minimum that corresponds to $AB$ stacking. The 
highest point of the barrier is $\sim65$ meV. This point is also global maximum 
where the S atoms from top and bottom layer are aligned on top of each other. 
This energy barrier is very small as compared to that of MoS$_{2}$ ($\sim200$ 
meV).\cite{Tongay3} In the zig-zag direction, the barrier profile results in a 
symmetric peak with the highest point obtained when the S atoms of the top and 
bottom layers are closest to each other at the path of sliding. The maximum 
value of the barrier  is $\sim50$ meV. This type of barrier is common for the T 
phase of TMDs. Following subsections investigate how these barrier profiles 
are modified by electric field, charging and loading pressure.

\begin{figure}[htbp]
\includegraphics[width=8.5 cm]{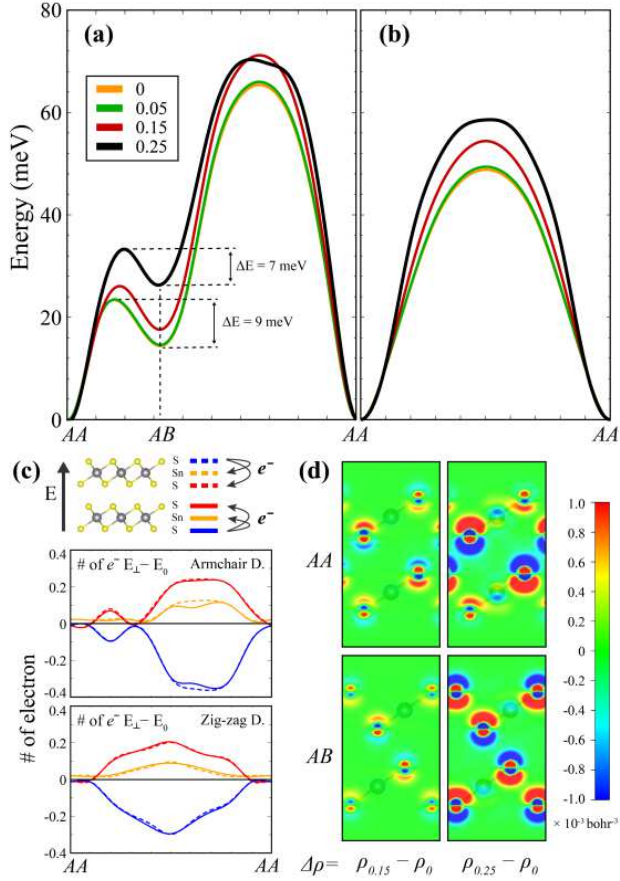}
\caption{\label{efield_fig}
(Color online) The plot of the energy barrier in the armchair (a) and the 
zig-zag (b) direction under zero and increasing electric
fields. The blue line 
represents the zero electric field case. The black, green and red lines are for 
0.05, 0.15, 0.25 V/\AA{} electric field cases, respectively. (c) electron transfer from outer S atoms to the inner 
part (upper) and the total number of electron difference (lower) 
between the 0.25 V/\AA{} electric field case and the zero electric field case 
on the sliding paths. (d) is the cross section of the total charge density 
difference between the 0.15, 0.25 V/\AA{} electric field cases and the zero 
field case for both $AA$ and $AB$ stacking orders. The color code of the 
isosurface values are 
given.}
\end{figure}

\begin{figure}[htbp]
\includegraphics[width=8.5 cm]{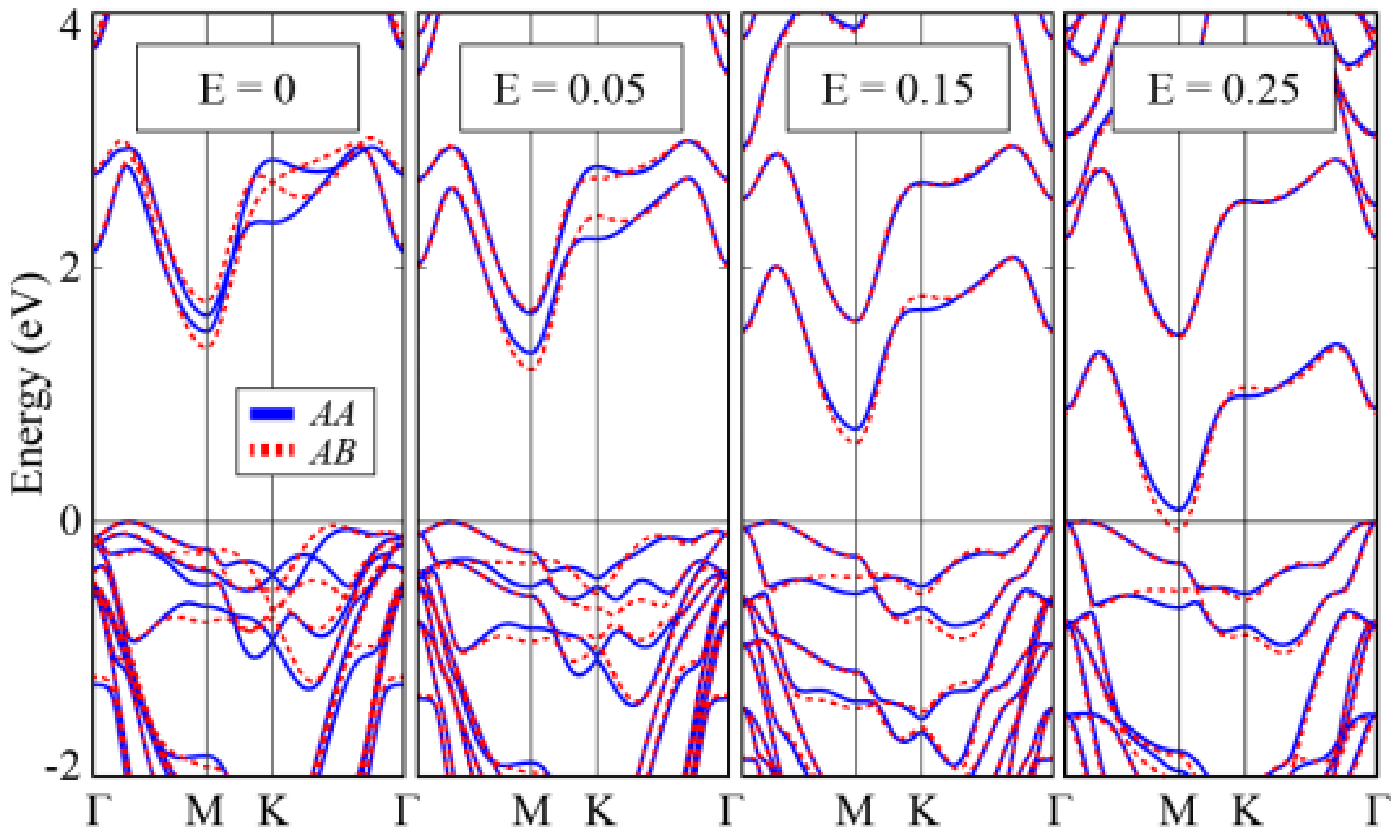}
\caption{(Color online) Energy band dispersions under three different E-field 
which are 0.05, 0.15, and 0.25 V/\AA{} perpendicular to the plane of structure. 
The blue curves are for $AA$ stacking and red curves are for $AB$ stacking.}
\label{bands_efield}
\end{figure}

\subsection{Effect of External Electric Field}\label{efield}

In this part, we investigate how the sliding potential is modified under the 
influence of 
an external perpendicular electric field (E-field). Three 
different (0.05, 0.15, 0.25 V/\AA) E-field values in the positive $z$-direction 
(perpendicular to the plane of bilayer system) are applied. While the energy 
barrier is conserved, the barrier heights 
increases as shown in Figs.~\ref{efield_fig} (a) and (b). The changes at
0.05 V/\AA{} field is negligible and the profiles are almost the same as in
the 
zero E-field case. In addition, for $AB$ stacking the energy difference 
between the local minimum and the neighboring local maximum decreases with 
applied E-field. The reason of the changes can be elucidated by analyzing the 
variations of the charge separation in the system. 

In Fig.~\ref{efield_fig} 
(c), by using the Bader charge analysis technique, the amount of charge 
difference on the atoms between the 0.25 V/\AA{} case and the zero field case 
is shown along the sliding path. The solid (dashed) orange curve represents the 
Sn atoms at the bottom (top) layer. The solid (dashed) blue curve represents 
the outer S atom at the bottom (top) layer and the solid (dashed) red curve is 
for the inner S atom (S atoms between the sublayers of Sn atoms). The charge 
configurations seem to be sensitive not only to the E-field but also the 
stacking order of the layers. Firstly, the charge variations of the Sn atoms 
of the bottom and the top layers are positive which indicates that the E-field 
shifts electron around the Sn atoms. On the other hand, the behaviors of the 
changes on the S atoms are different according being at the outer or the inner 
part of the bilayer system. The outer S atoms have less electrons under E-field 
while inner S atoms attain more electrons. As an exception to these trends, the 
charge of the S atoms is not altered much by the E-field for the $AA$ and the 
$AB$ stackings. It needs more detailed analysis.

To clarify the effect of the external E-field on the $AA$ and 
$AB$ bilayer systems, the total charge density difference between with- and 
without-E-field for the cross section 
through the atoms in the unitcell are shown in Fig.~\ref{efield_fig}(d). It 
seems that the S atoms are polarized by the E-field, but the Sn atoms experience 
no significant change. In the case of 0.15 V/\AA{}, the polarizations are larger 
at the inner sides of S atoms for both $AA$ and $AB$ stackings. In the case of 
0.25 V/\AA{}, the polarization vanishes at the outer S atoms for $AA$ stacking 
order. On the 
other hand, the polarization still exists and is enhanced at the inner S atoms 
of $AA$ and all S atoms of $AB$ stacking. More importantly, the number of 
electron increases gradually between the layers with electric field strength 
for both the $AA$ and the $AB$ stackings. This charge accumulation between the 
layers is consistent with the study of Ramasubramaniam \textit{et 
al.}\cite{Ramasubramaniam} where the MoS$_{2}$ bilayer is tuned by the external 
E-field and the charge distribution between the layers was gradually enhanced 
with increasing out of plane E-field. According to our results, the 
perpendicular E-field increases the coupling between the SnS$_{2}$ layers 
for $AA$ stacking as compared to $AB$. 

In addition, the 
E-field dramaticaly modifies the electronic structure of the bilayer system as 
shown in Fig.~\ref{bands_efield}. Under the E-field, the VBM approaches to 
$\Gamma$ point while the CBM at $M$ point drops in energy which means that band 
gap decreases. The drop of band gap with E-field is slower for the $AA$ 
stacking, so under $0.25$ V/\AA{} E-field, the $AB$ become semi-metal while the 
$AA$ stacked bilayer is semiconductor with band gap 85 meV within GGA. Although 
this is the underestimated band gap, the trend of change on electronic 
structure together with enhanced stacking strength at $AA$ order indicates that 
perpendicular E-field is useful method for tunning the band gap of bilayer 
SnS$_{2}$ which is needed in a material for the field effect transistor 
application.

\begin{figure}[htbp]
\includegraphics[width=8 cm]{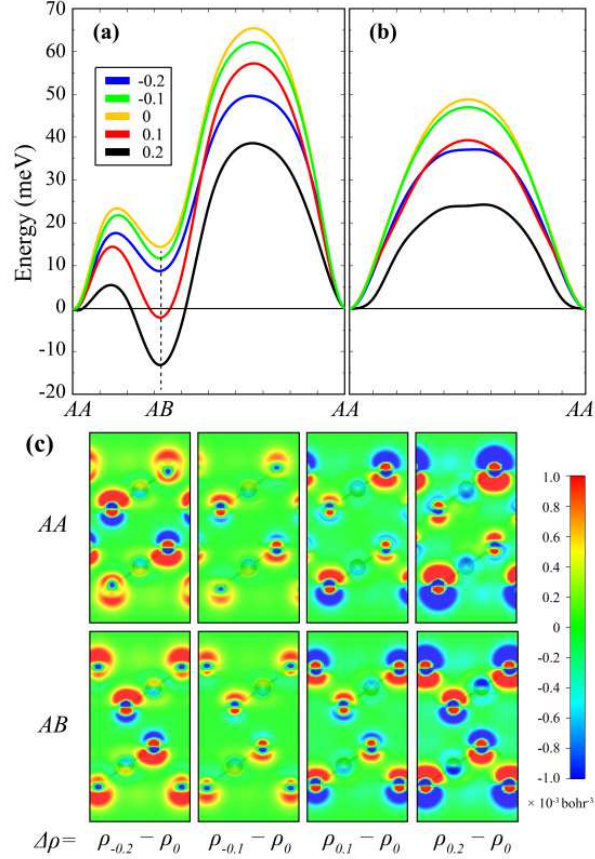}
\caption{\label{charging_fig}
(Color online) Plot of the energy barrier for sliding in the armchair (a) and 
the zig-zag (b) direction for four different charging cases and the bare case. 
Positive value of charing refers to a higher electron density. The blue curves 
represents the bare case. The green, the black, the yellow and the red curves 
are for the charging case of $-0.2$, $-0.1$, $0.1$, $0.2$ $e^{-}$. (c) Cross 
section of the total charge density difference between the charged cases and 
the bare case.}
\end{figure}

\subsection{Effect of Charging}\label{charging}

The sliding-energy barrier can also be tuned by 
controlling the total charge on 
the system. In this part, we examine the modifications of the sliding barrier 
by adding or subtracting electron (doping electron or hole) to the bilayer. In 
Figs. \ref{charging_fig} (a) and (b), we show respectively the barriers forms 
along the armchair and the zig-zag directions for four different charging 
conditions. Positive value of charging refers to extra electrons. First of all, 
it is interesting that the 0.2 $e^{-}$ and also the 0.1 $e^{-}$ (per unit cell) 
cases result in a minimum energy for the $AB$ stacking instead of $AA$. In 
addition, the barrier height decreases in all charging conditions and the shape 
of the barrier differs considerably for the armchair direction. Moreover, the 
maximum barrier height in the zig-zag direction decreases down to $\sim$20 meV 
which is comparable with the thermal energy at room temperature (25 meV). 

The effect of charging can be understood by monitoring the charge 
localizations for $AA$ and $AB$ stacking. 
Therefore, in Fig.~\ref{charging_fig}(c), the cross section of the charge 
density differences between the charged and the bare systems is shown for $AA$ 
and $AB$ stackings. It is expected that only positively charged regions to 
appear in the plot for the electron subtracted system. For the electron-added 
system, the expectation is the opposite. However the occurrence of both 
positive and negative regions for each charging case indicates that charging 
(positive or negative) modifies the distribution of the other electrons. In 
Fig.~\ref{charging_fig}(c), it is explicitly seen that the inserted charges 
accumulate to the outer surfaces (red for positive charges and blue for the 
negative charges) of the bilayer system. In all cases, except the 0.2 $e^{-}$ 
added to $AA$ stacking, the electrons (blue region) also accumulate to the 
region between the layers. When we compare the $AA$ and $AB$ stackings in 
all charging conditions, the lower energy case has always a larger amount of 
electrons (negative charge means blue region) between the layers
which indicates that the interactions between the layers 
have covalent character. 

\begin{figure}[htbp]
\includegraphics[width=8.5 cm]{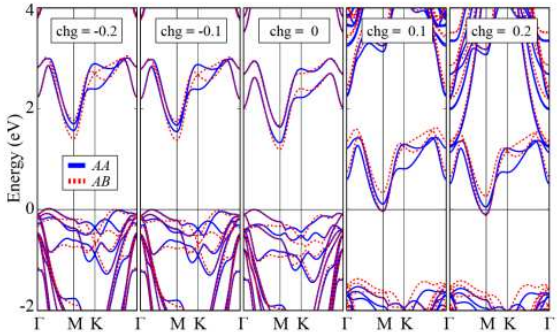}
\caption{(Color online) Energy band dispersions for four different charging 
conditions which are $-0.2$, $-0.1$, $0.1$, and $0.2$ $e^{-}$. The negative 
numbers refer the electron subtracted (hole doping) cases and the positive 
numbers refer the electron added (electron doping) cases. The blue curves are 
for $AA$ stacking and red curves are for $AB$ stacking.}
\label{bands_chg}
\end{figure}

In addition, when electrons are doped, $s$ orbital of 
Sn atoms which mostly construct the CBM are firstly occupied as shown in 
Fig.~\ref{bands_chg}. The energy difference between the newly occupied Sn states 
and the already occupied S states decreases. The change is higher for the $AB$ 
stacking as compared to $AA$. For the hole doping case, $p_{x}$ and $p_{y}$ 
orbitals of S atoms which are dominant around the Fermi level (VBM) are firstly 
occupied. To sum up, both electron and hole doping decreases sliding barrier 
which makes easier to modify the stacking order, and for proper value of 
electron doping, favorable stacking order become $AB$ stacking instead 
of $AA$.

\subsection{Bilayer Under Loading Pressure}\label{pressure}

The energy landscape of bilayers under constant loading pressure is studied for 
various 2D materials.\cite{Friction} It is shown that for a given value of the 
applied pressure, the ratio between intralayer and interlayer interaction is a 
material property that describes the transition from the stick-slip to the 
superlubric regime. Here we study the effect of a constant loading pressure on 
the energy landscape of bilayer SnS$_2$.

\begin{figure}[htbp]
\includegraphics[width=8.5 cm]{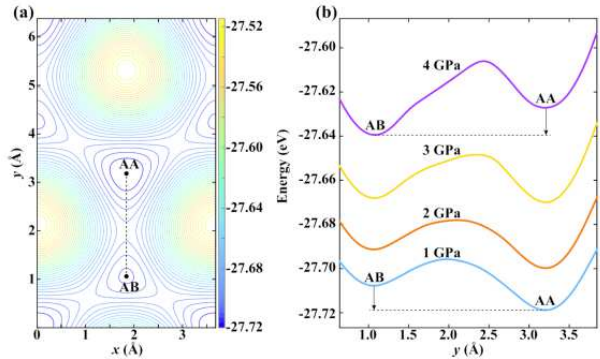}
\caption{(Color online) (a) Constant height energy landscape of bilayer 
SnS$_2$. 
The height is fixed at the value corresponding to the distance between the top 
and the bottom sulfur atoms when the bilayer is fully relaxed. (b) The 
constant pressure energy dependence along the line connecting the $AA$ and 
the $AB$ stackings.}
\label{pressure_fig}
\end{figure}

The $AA$ and the $AB$ stackings of bilayers, composed of materials like 
MoS$_2$, have the same energy due to the symmetry of the 1H structure. However, 
the energy of bilayer SnS$_2$ in 1T is different for $AA$ and $AB$ stackings. 
This is evident from the previous figures in this section as well as 
from Fig.~\ref{pressure_fig}(a) where we present the constant height energy 
landscape of the SnS$_2$ bilayer. To calculate the energy landscape at constant 
pressure we repeat constant height scans by lowering the height by 0.2 \AA{} 
steps. In this way we get the energy for the three-dimensional movement of the 
layers with respect to each other. We use this data to create the plots 
presented in Fig.~\ref{pressure_fig}(b). Here for a chosen loading pressure we 
first find the corresponding force in the $z$-direction. Then using spline 
interpolation we calculate the height that gives this force for each position 
in the $xy$ plane while moving from $AA$ stacking to $AB$ stacking. 
Interestingly, as the applied pressure is increased the relative energy 
difference between $AA$ and $AB$ stacking decreases and become even zero 
at 3 GPa. For pressures exceeding 3 GPa $AB$ stacking becomes more 
favorable than $AA$ stacking.

\section{Conclusion}\label{conc}

Starting from the monolayer, the electronic and the optical properties of 
bilayer SnS$_{2}$ are investigated within first principles DFT calculations. We 
found that the interaction between the layers is weaker than that of MoS$_{2}$ 
and other common TMDs. We showed that although the layers interact weakly the 
energy band  gaps and the absorbance spectra could be informative about the 
stacking type of the bilayer system. The energy barrier for the sliding of one 
layer over the other is found to be $\sim65$ meV at its maximum which is also 
small as compared to MoS$_{2}$.

The effect of applied E-field, charging and loading pressure on the sliding 
barrier of bilayer SnS$_2$ were also studied. Under the 
influence of a perpendicular E-field, for the $AA$ stacking which is favorable 
for the bilayer system, the coupling of the layer strengthens and consequently 
the sliding barrier height increases. In addition, it is shown that band 
gap of the 
bilayer SnS$_{2}$ can be tuned by perpendicular E-field and under 
sufficient E-field it can be turned from semiconductor to semi-metal. On the 
other hand, both adding and subtracting electrons decreases the barrier. More 
significantly, under charging or loading pressure, $AB$ stacking order can 
become the favorable configuration instead of $AA$ stacking. Tunable 
bandgap makes 2D crystal of SnS$_{2}$ a promising 
material for nanometer size field effect transistor applications. Furthermore, 
due to its easy-tunable stacking sequence, layered SnS$_{2}$ is also a good 
candidate for nanoscale lubricant applications.

\section{Acknowledgments} 

This work was supported by the bilateral project between TUBITAK (through Grant 
No. 113T050) and the Flemish Science Foundation (FWO-Vl). The calculations were 
performed at TUBITAK ULAKBIM, High Performance and Grid Computing Center 
(TR-Grid e-Infrastructure). CB, HS, and RTS acknowledge the support from 
TUBITAK Project No 114F397. H.S. is supported by a FWO Pegasus Marie Curie 
Fellowship. SC and AR acknowledges financial support from the Marie Curie 
grant FP7-PEOPLE-2013-IEF Project No. 628876, the European Research Council 
(ERC-2010-AdG-267374), Spanish grant, Grupos Consolidados (IT578-13). SC 
acknowledges support from The Scientific and Technological Research Council of 
Turkey (TUBITAK) under the project number 115F388.

\end{document}